\documentclass[twocolumn,showpacs,amsmath,superscriptaddress,amssymb]{revtex4}

\usepackage{graphicx}
\usepackage{amssymb}
\usepackage{color}

\begin{document}

\newcommand{\be}{\begin{equation}}
\newcommand{\ee}{\end{equation}}
\newcommand{\bea}{\begin{eqnarray}}
\newcommand{\eea}{\end{eqnarray}}
\newcommand{\rv}{{\bf r}}
\newcommand{\nv}{\hat{{\bf n}}}
\newcommand{\bv}{\hat{{\bf b}}}
\newcommand{\ev}{\hat{{\bf e}}}
\newcommand{\xv}{{\bf x}}
\newcommand{\kv}{{\bf k}}
\newcommand{\qv}{{\bf q}}
\newcommand{\Gv}{{\bf G}}
\newcommand{\uv}{{\bf u}}
\newcommand{\sig}{\sigma}
\newcommand{\Rv}{{\bf R}}
\newcommand{\tv}{\hat{{\bf t}}}
\newcommand{\Tv}{\hat{{\bf T}}}
\newcommand{\Nv}{\hat{{\bf N}}}
\newcommand{\Bv}{\hat{{\bf B}}}
\newcommand{\cH}{{\cal H}}
\newcommand{\cT}{{\cal T}}
\newcommand{\cK}{{\cal K}}
\newcommand{\grad}{{\bf \nabla}}

\title{Mesophases of soft-sphere aggregates}

\author{Homin Shin}
\affiliation{Department of Polymer Science and Engineering,
University of Massachusetts, Amherst, MA 01003, USA}
\author{Gregory M. Grason}
\affiliation{Department of Polymer Science and Engineering,
University of Massachusetts, Amherst, MA 01003, USA}
\author{Christian D. Santangelo}
\affiliation{Department of Physics,
University of Massachusetts, Amherst, MA 01003, USA}

\begin{abstract}
Soft spheres interacting via a hard core and range of attractive and repulsive ``soft-shoulder" potentials self-assemble into clusters forming a variety of mesophases.   We combine a mean field theory developed from a lattice model with a level surface analysis of the periodic structures of soft-sphere aggregates to study stable morphologies for all clustering potentials.  We develop a systematic approach to the thermodynamics of mesophase assembly in the low-temperature, strong-segregation and predict a generic sequence of phases including lamella, hexagonal-columnar and body-center cubic phases, as well as the associated inverse structures.  We discuss the finite-temperature corrections to strong segregation theory in terms of Sommerfeld-like expansion and how these corrections affect the thermodynamic stability of bicontinuous mesophase structures, such as gyroid.  Finally, we explore the opposite limit of weakly-segregated particles, and predict the generic stability of a bicontinuous cluster morphology within the mean-field phase diagram.
\end{abstract}


\maketitle

\section{Introduction}

New classes of soft materials assemble into a bewildering array of complex mesophases. These include bicontinuous structures with multiple symmetries \cite{cochran_bates}, unusual lattices such as the A15 \cite{percec, ziherl, grason_didonna}, as well as the more exotic Frank-Kasper and quasicrystalline packings of dendrimers and block copolymers \cite{zeng04,ungar05,hayashida07}. This rich behavior holds out the promise of designing materials to achieve a complex target structure on length scales long enough to be used as optical devices or as components of hybrid-photovoltaic structures.  Nevertheless, the principles connecting structure and interactions at the molecular scale to the structure and thermodynamics of long-range assemblies remain poorly understood.

In this article, we study the stable morphologies of a class of interacting particles with a hard core and an isotropic, soft corona. The class is exemplified by a corona which costs a fixed energy $\epsilon$ when it overlaps with the corona of another particle.  At sufficient densities, this interaction results in stable, finite size clusters arranged into a number of structures \cite{likos01,likos, jagla, malescio, glaser_et_al,kahl,kahl2,kahl3,kawakatsu}. In two dimensions, these clusters arrange into either hexagonally-packed clusters, stripes or a dense phase of hexagonally-packed voids \cite{jagla, malescio, glaser_et_al,kahl,frenkel08}. In three dimensions, numerical calculations of the ground state show a number of distinct lamellar, cylindrical and spherical clusters as well as their inverse form-like structures \cite{kahl, kahl2, kahl3}. This behavior is by no means specialized, however, as a range of particle interaction types have demonstrated to mesophase clustering. Another classic example of cluster morphologies occurs for potentials with long-range repulsion and short-range attractions \cite{gelbart,archer07, coniglio06}. Cluster phases have also been demonstrated with magnetic dipoles in a magnetic field in two dimensions both in simulations \cite{camp} and, more recently, in experiments \cite{ziherl07}.

Beyond isotropic particle systems, it is striking that predicted cluster morphologies bear a strong resemblance to the phase morphologies of block copolymer melts~\cite{bates_fredrickson_arpc}.  Despite the obvious differences in the microscopic structure of polymeric amphiphiles and isotropic particles with hard cores, we find that the gross features of the long-range order expected in both systems to be in exact correspondence both in terms of topology of aggregate structures as well as the symmetry of their periodic arrangement.

In this paper we explore the full potential of isotropic, clustering particle systems to form long-range ordered mesophases and investigate the thermodynamic principles underlying their assembly.  In further analogy to the theory of block copolymer thermodynamics, we develop a mean-field approach to clustering morphologies which is tractable both in the limit of strong-segregation~\cite{semenov}, valid at low temperature and sharp cluster boundaries, and weak-segregation~\cite{leibler}, valid near the order-disorder transition.  In addition to predicting the ground state morphologies, our low-temperature expansion allows to probe the explicit relationship between particle entropy, interactions and cluster geometry.  This expansion highlights a frustration between the particle interaction potentials and suggests a mechanism whereby cluster morphologies with high volumetric surface areas are stabilized at higher temperatures. Interestingly, we predict a stable orthorhombic, {\it Fddd} bicontinuous phase in a narrow region near the order-disorder transition in complete analogy to the prediction \cite{tyler_morse, ranjan_morse} and observation\cite{cochran_bates, epps_cochran_bates} of this same structure in copolymer melts.

Previous studies of clustering from isotropic particles have largely focussed on properties of a specific type of interaction potential. However, in ref. \cite{likos} a more general classification of interaction potentials is introduced. Here, we take this one step further, demonstrating that the gross properties of equilibrium assembly are rather insensitive to the precise details of the potential. This universality stems largely from universal geometric considerations both at high temperatures, where this is expected \cite{seul95}, and at low temperatures.

In section \ref{sec:theory}, we discuss the general features of soft assemblies from the point of view of the Fourier transform of the interaction potential. In section \ref{sec:lowT}, we derive the phase diagram using a strong-segregation theory, valid when the cluster interfaces are sharp and the temperature is small. In section \ref{sec:finiteT}, we expand on these results with a finite temperature expansion. Finally in section \ref{sec:highT}, we present a phase diagram valid near the order-disorder transition. In section \ref{sec:discussion}, we summarize our results. We also discuss a paradigm for designing soft matter clusters.

\section{Statistical Mechanics of Cluster Mesophases}\label{sec:theory}
\subsection{Clustering instability and particle interactions}

We begin our analysis by dividing the interaction between particles into two parts, the hard-core $V_{HC}(r)$ potential, where
\begin{equation}
V_{HC} (r) = \left\{ \begin{array}{ll} \infty,&  r<a \\
0, & r> a \end{array} \right.
\end{equation}
and ``soft-shoulder" $V_{SS}(r)$ which characterize the longer range behavior of particle interactions.  On its own, the ground states of the step-function corona described in the introduction
\begin{equation}
\label{eq:step}
V_{SS} (r) = \left\{ \begin{array}{ll} \epsilon, &  r<\sigma  \\
0, & r> \sigma \end{array} \right.
\end{equation}
solve a particularly simple optimization problem.  Since this potential is purely repulsive, the different morphologies that are observed, apparently, minimize the number of overlaps within a distance less than $\sigma$  at fixed density \cite{kahl2}.
It is not clear, however, how other types of clustering systems can be explained with such a simple geometrical optimization. Moreover, designing clustering particles that exploit `soft packing' requires an additional understanding of what principles guide cluster morphology.

\begin{figure}[t] \centering \includegraphics[angle=0, scale=0.5]{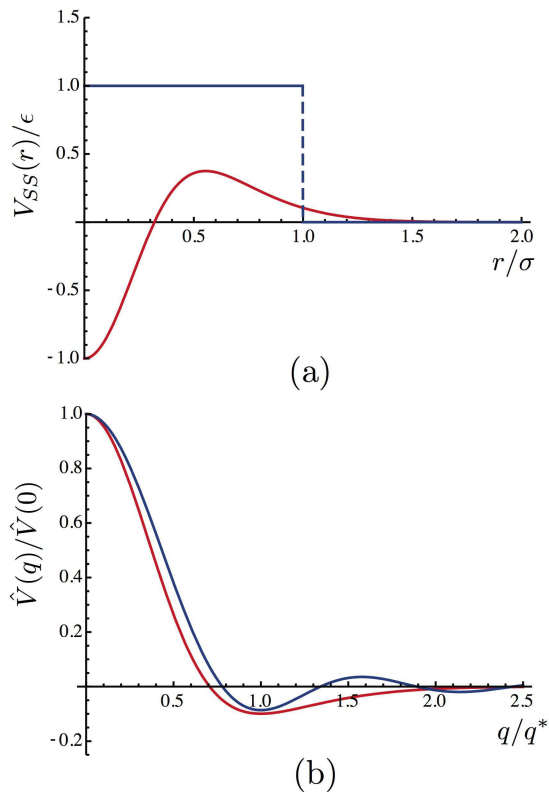} \caption{(Color online) Two real-space interparticle soft-should potentials $V_{SS}(r)$ are shown in (a): eq. (\ref{eq:step}) in blue and eq. (\ref{eq: lrrsra}) in red with $A=2 \epsilon$, $B=\epsilon$, $\lambda_A=\sigma/3$ and $\lambda_B=2\sigma/3$.  In (b) the Fourier transforms of each potential, normalized by their respective values at $q=0$.  Both the long-range repulsive/short-range attractive and step repulsive potential exhibit a minimum for which $\hat{V}(q^*)<0$, indicating an instability to clustering at sufficiently low temperatures.} \label{potentials} \end{figure}

By examining the soft-shoulder potential from the point of view of its Fourier transform,
\begin{equation}
\hat{V}({\bf q}) = \int d^3 r V_{SS}( \rv) e^{i {\bf q} \cdot \rv } ,
\end{equation}
the more general conditions for the clustering behavior become apparent.  In refs. \cite{likos} and \cite{glaser_et_al}, the authors observed that, at least within a mean-field approximation, interaction potentials whose Fourier transforms have a negative region exhibit an instability leading to clustering. Precisely, the instability occurs when $1 + \beta \hat{V}({\bf q}) \rho(1-\rho/\rho_m) < 0$, where $\rho_m$ is the maximum packing density of cores, $\beta = 1/k_B T$, $\rho$ is the average density \cite{glaser_et_al}.

It is, perhaps, not surprising that certain combinations of attractive and repulsive interactions can satisfy this condition. For example, consider the interaction potential
\begin{equation}
\label{eq: lrrsra}
V_{SS} (r) = -A \exp \left(-r^2/\lambda_A^2\right) + B \exp\left(-r^2/\lambda_B^2\right).
\end{equation}
where $A$ and $B$ are positive and $\lambda_A/\lambda_B < 1$. Therefore, at distances $r > \lambda_B$ the soft-shoulder potential is repulsive but when $r < \lambda_A$ it is attractive.
The Fourier transform is
\begin{equation}
\hat{V} ({\bf q}) = - \frac{A \lambda_A^{3}}{2 \sqrt{2}} \exp \left(-\lambda_A^2 q^2/4 \right) + \frac{B \lambda_B^{3}}{2 \sqrt{2}} \exp \left(-\lambda_B^2 q^2/4 \right),
\end{equation}
which has a single negative peak at a non-zero wave vector, provided that $A \lambda_A^{3} < B \lambda_B^{3}$.   More interesting is that purely repulsive potentials, like the step potential in eq. (\ref{eq:step}), can exhibit the same negative peak.  Figure \ref{potentials} shows that these two very different soft shoulder forms give rise to a similar properties of $\hat{V} ({\bf q})$ in Fourier space, namely a minimum at some finite wavevector $|{\bf q}|=q^*$ for which $\hat{V} ({\bf q}^*)<0$.  Not even sharp discontinuities are required: simulations \cite{camp} and experiments \cite{ziherl07} in two-dimensions of a dipolar magnetic repulsion softened by an attractive component show clustering behavior even in regimes in which the total potential is repulsive.

From this point of view the fundamental frustration occurring in clustering particle systems becomes transparent.  The soft shoulder portion of the interaction potential drives the system to order at a particular wavelength, $\lambda_*=2\pi/q^*$, associated with minimum in $\hat{V} ({\bf q})$.  At the onset of order, this may be accomplished by a weak sinusoidal modulation of the density profile.  At lower temperatures, the soft-shoulder interaction prefers a much stronger density modulation, favoring particle packings that are at higher densities than can be accommodated by the hard-core of the particle.  Hence, the hard-core portion of the interaction prohibits ``over-crowding" of particles, forcing aggregates to spread out into clusters of various morphologies.

This observation provides an alternate methodology for soft sphere packing problems involving the features of the Fourier transform of the potential. Moreover, similarities in the Fourier transform of two potentials of very different origins lead one to an effective optimization problem expressed entirely in terms of minimizing the number of corona overlaps subject to constraints on the particle density.

In further sections of this article, we will focus on potentials whose Fourier transforms--like those illustrated in Fig. (\ref{potentials})--are dominated by two modes:  the modes at ${\bf q}=0$ and $|{\bf q}|=q^*$.  This requires that $|\hat{V} ({\bf q}) /\hat{V} ({\bf q}^*) |\ll 1$, for wave numbers at other negative peaks ${\bf q} \neq {\bf q}^*$, as we find for the potentials discussed in this section~\footnote{In addition, we focus on the limiting case where the peak width $\Delta q^*/q^* \ll 1$ -- the negative peak should not be too shallow.}.  Additionally, we assume that $\hat{V}({\bf q})$ is maximum  and finite at ${\bf q}=0$.

\begin{table*}[ht]
\centering
\begin{tabular}{c c c c c }
\hline
phases & ${\bf G}$ & \,\,\,m \,\,\, &$\psi({\bf x})$ & ranges of level set  $\psi_0$   \\
\hline \hline
lamellar &  $\pm \hat x$  & 2&$2\cos(x)$    &   $-2<\psi_0<2$   \\
hexagonal &  $\pm \hat x, \pm \frac{1}{2}\hat x \pm \frac{\sqrt 3}{2} \hat y$  & 6 &$2\left[ \cos(x) + 2\cos(\frac{x}{2})\cos(\frac{\sqrt 3}{2}y)\right] $  &   $-2<\psi_0<6$  \\
sc or P surface & $\pm \hat x, \pm \hat y, \pm \hat z$ & 6&$2\, [\cos(x)+\cos(y)+\cos(z)]$  & $-6<\psi_0<6$  \\
bcc or IWP surface \,\,& $ \pm \hat x \pm \hat y, \pm \hat y \pm \hat z, \pm \hat z\pm \hat x$ & 12& \, \, $4\, [\cos(x)\cos(y)+\cos(y)\cos(z)+\cos(x)\cos(z)]$ \, \,&  $-4<\psi_0<12$  \\
fcc & $\pm \hat x \pm \hat y \pm \hat z$ &8 &$8\, [\cos(x)\cos(y)\cos(z)]$ &     $-8<\psi_0<8$    \\
gyroid & $\pm \hat x \pm \hat y, \pm \hat y \pm \hat z, \pm \hat z\pm \hat x$&12 &$4\,[\sin(y)\cos(z)+\sin(z)\cos(x)+\sin(x)\cos(y)]$  &  $-6<\psi_0<6$
\\\hline
  \end{tabular}
  \caption{Level surface functions for various phases. $\bf G$ is the primary reciprocal vector and
  $m$ is the number of modes for $|{\bf G}|=q^*$. Notice that the gyroid phase is constructed by a phase shift $\delta_{\Gv}=\pi/2$.}
  \label{levelfunction}
\end{table*}

\subsection{Lattice model particle clustering}

\begin{figure*}[ht] \centering \includegraphics[angle=-90, scale=0.60]{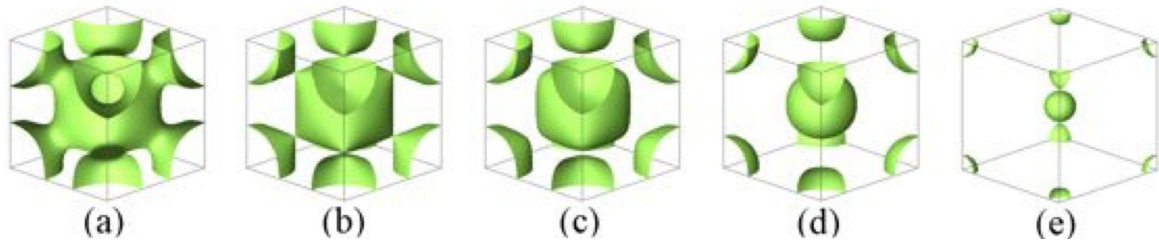} \caption{(Color online) Level surface
morphologies in terms of level set constants $\psi_0$ for $Im{\bar3}m$ space group. I-WP surface with $\psi_0=1.0$ (a) starts
to pinch off at $\psi_0=0.0$ (b) and evolves to bcc phases $\psi_0=0.4$ (c), $\psi_0=4.0$ (d), and $\psi_0=10.0$ (e).}
\label{IWP} \end{figure*}

To strip away the complexity arising from the hard sphere packing of the particles, we follow ref. \cite{glaser_et_al} and introduce a lattice gas model to describe the mesophase behavior of hard core/soft shoulder particles.  In order to prevent overlaps of particle cores, this model assumes particles to occupy a close-packed lattice of cores.   Though this type of model only approximates the translational entropy of particle cores, we expect the model to become accurate in the low temperature regime where clusters are densely packed and when core size is much smaller than dimensions of the clusters, $\lambda_* \gg a$.  The state of each lattice site $i$ is represented by $n_i = 0$ or $1$. The soft-shoulder pair interaction between sites $i$ and $j$ is represented by $V^{i j}_{SS}=V_{SS} (|{\bf x}_i-{\bf x}_j|)$. The effective Hamiltonian is then written as
\be \label{lattice}
\mathcal{H}[n_i]=\frac{1}{2} \sum_{ij} n_iV^{ij}_{SS}n_j -\sum_i
\mu n_i  \ , \ee
where $\mu$ is the chemical potential that controls the density of particles in the system.
After a delicate Hubbard-Stratonovich transformation, the grand partition function can be written in terms of a coarse grained field $\phi({\bf x})$~\cite{glaser_et_al} by
${\cal Z} = \int \mathcal{D} \phi ({\bf x}) \exp \,  \{-A[\phi({\bf x})]\}$,
where the action $A$ is
\bea \label{action}
A [\phi({\bf x})]&=& \frac{1 }{2\beta } \int d^3 x \int d^3 x' \,
\phi({\bf x}) V_{SS}^{-1}({\bf x}-{\bf x'})\phi({\bf x'}) \nonumber \\
&-&\rho_m \int d^3 x \, \ln \, [ 1+ \exp (\beta \mu +i \phi({\bf x})) ] \ ,
\eea
where,%%Check this next equation!!
\begin{equation}
V^{-1}_{SS} ({\bf x}) = \int \frac{ d^3 q}{(2 \pi)^3} \frac{ e^{i {\bf q} \cdot {\bf x} } }{\hat {V} ({\bf q}) } .
\end{equation}
The convergence of the path integral requires that $\phi$ be purely real for Fourier modes where $\hat{V}(\qv)>0$ and purely imaginary for modes where $\hat{V}(\qv)<0$~\cite{fisher_park}.  The mean-field analysis we pursue here is insensitive to this technicality.

We pursue our analysis of eq. (\ref{action}) by making two important approximations. First, we specialize to the mean-field limit, in which $\phi(\textbf{x})$ is given by the critical point of $A[\phi(\textbf{x})]$.
Minimizing the action with respect to $\phi({\bf x})$ leads to
the mean-field equation, for each Fourier component,
\be \label{mean_s}
 \hat{\phi} ({\bf q}) = i \beta \hat{V} ({\bf q} ) \hat{\rho} ({\bf q}) \ ,
\ee
where $\hat{\rho}_{\bf q} = \mathcal{V}^{-1} \int d^3x~ \rho({\bf x}) \exp(-i{\bf q}\cdot {\bf x})$ and $\mathcal {V}$ is the total volume.  Along with the mean-field condition we also have the relation between $\phi(\xv)$ and mean particle density for the lattice gas model
\be
\label{rho}
\rho(\xv_i) = \rho_m \langle n_i \rangle =   \rho_m \frac{e^{\beta \mu + i \phi(\xv_i)}}{ 1+ e^{\beta \mu + i \phi(\xv_i)} } .
\ee
Together eqs. (\ref{mean_s}) and (\ref{rho}) form a self-consistent set of equations for the mean-field density profile, $\rho(\xv)$, and self-consistent field, $\phi(\xv)$.

The validity of this saddle-point approximation is determined by strength of interparticle interactions relative to the thermal energy scale.  Assuming that mean-field solutions $\phi \propto i \beta \hat{V} (\qv =0) $  and the mean-field action becomes linear in $\beta \hat{V} (0)  \rho_m^2 {\cal V}$. The mean-field analysis, therefore, requires a steep saddle, or sufficiently strong particle interactions $\beta \hat{V} (0) \rho_m \gg 1$.

\subsection{Level set representation of mean-field solutions}

Clustering behavior is observed when the Fourier transformed potential $\hat V({\bf q})$ takes on negative values for some ${\bf q}$.  This suggests that equilibrium field configurations will be periodic structures whose fundamental mode will have length $|{\bf q}| = q^*$.  Further, from the mean-field condition eq. (\ref{mean_s}), we see that if $\hat{V} ({\bf q})$ is dominated by $q=0$ and $q=q^*$ modes, we may consider configurations of $\phi({\bf x})$ that contain only these wavevectors.  Note that this applies equally to the case of strong-segregation where the mean-field density has many non-zero Fourier modes beyond its fundamental mode.  Thus, for a given periodic structure, characterized by a set of reciprocal lattice vectors, ${\bf G}$, we write $\phi({\bf x})$ as a superposition of plane waves with only $q=0$ and $q=q^*$ modes:
\be\label{eq:expansion}
\phi ({\bf x}) = \hat{\phi}_0 + \sum_{|{\bf G}| = q^*} \hat{\phi}_{\bf G} e^{i {\bf G}\cdot {\bf x} } \ .
\ee
For a given symmetry group, we can take advantage of the fact that Fourier transforms along different reciprocal lattice directions have the same amplitude to rewrite field solutions as
\be
\phi({\bf x}) - \hat{\phi}_0 = \hat{\phi}_{{\bf G}} \psi(\xv)
\ee
where
\be
\label{level}
\psi(\xv) = \sum_{|\Gv|=q^*} e^{i {\bf G}\cdot {\bf x} + i \delta_\Gv } ,
\ee
where $\delta_\Gv$ are mode-dependent phases.  For a given symmetry, all information about the three-dimensional, equilibrium cluster morphology is encoded in $\psi(\xv)$, the {\it level set function}.

Using the mean-field values of $\hat{\psi}_0$ and $\hat{\psi}_\Gv$ from eq. (\ref{mean_s}) and defining,
\begin{equation}
\psi_0 \equiv \frac{ \hat{V}_0\hat{\rho}_0 \ -\mu} { \hat{\rho}_* | \hat{V}_*|}
\end{equation}
and
\begin{equation}
\alpha \equiv \beta   | \hat{V}_*| \hat{\rho}_* ,
\end{equation}
we may write the mean-field solution form of the density as,
\begin{equation}
\label{rholevel}
\rho(\xv)/\rho_m = \frac{ e^{\alpha(\psi(\xv) - \psi_0)} }{1+ e^{\alpha(\psi(\xv) - \psi_0)} } .
\end{equation}
Here, we use the notation $\hat{V}_0 = \hat{V} (0)$,  $\hat{V}_* = \hat{V} (q^*)$, $\hat{\rho}_0 = \hat{\rho}(0)$ and $\hat{\rho}_* = \hat{\rho}(q^*)$ .  Regions of high (low) particle density correspond to regions where $\psi(\xv) > \psi_0$ ($\psi(\xv)< \psi_0$); the boundaries of the clusters are defined by the surfaces where $\psi(\xv) = \psi_0$.  Hence, adjusting the level-set parameter $\psi_0$ adjusts the size, structure and surface of aggregates~\cite{thomas}.

In table~\ref{levelfunction}, we list level set functions which correspond to single-mode plane-wave superpositions of the type defined in eq. (\ref{level}).  These include the one-dimensional lamellar structure and two-dimensional hexagonal columnar structure.  Three-dimensional structures include cubic arrays of spheres:  simple-cubic (sc); body-centered cubic (bcc); and face-centered cubic (fcc).  Additionally, these include multiply-connected bicontinuous structures, which we refer to by the name of the minimal surface of the same topology:  IWP surface; Schwarz's P surface, otherwise known as the ``plumbers nightmare"; and the gyroid (G) surface~\cite{thomas}.  In the representation of equation \ref{level}, two distinct morphologies can share the same level surface function $\psi({\bf x})$. For example, the bicontinuous I-WP surface, having $Im\bar{3}m$ space group, occurs in the parameter range $-4<\psi_0<0$, whereas the bcc structure occurs in $0<\psi_0<12$.
Similarly, Schwarz's P surface shares the same level function as the sc phases: sc phases occur in the range $2<|\psi_0|<6$ whereas the P surface occurs when $-2<\psi_0<2$ ($Pm\bar{3}m$ space group).
In Fig.~\ref{IWP}, we display a representative example of cluster shapes with increasing $\psi_0$ for the $Im\bar{3}m$ space group.

In terms of the mean-field solutions, themselves described by a family of level-set functions, we now have the grand potential in the saddle-point approximation,
\bea \label{grandp}
\frac{\beta\Omega}{{\cal V}} &=& \frac{\beta }{2} \left(
m|\hat{V}_*| \hat{\rho}_*^2 -\hat {V}_0 \hat{\rho}_0^2\right) \\ \nonumber
&-&\rho_m\int d^3 x \, \ln \, \left[ 1+ \exp \{ \alpha
(\psi({\bf x}) -\psi_0) \} \right] \ ,
\eea
where $m$ is the number of plane-wave modes in $\psi(\xv)$.

\section{Low temperature, strong-segregation limit}\label{sec:lowT}
In the limit of low temperatures the parameter $\alpha \gg 1$, indicating from eq. (\ref{rholevel}) that clusters are separated by sharply defined boundaries dividing regions of densely packed clusters with $\rho(\xv)  = \rho_m$ from empty regions with $\rho(\xv) =0$.  In absolute limit of $\alpha \to \infty$, the temperature dependence of all quantities is eliminated.  In particular, the value of the grand potential for a particular mean-field solution has the form,

\begin{multline}
\label{SST}
\frac{ \tilde \Omega (|\tilde{V}_*|, \tilde \mu)}{{\cal V}} = \frac{1}{2} \left(
m|\tilde{V}_*| \hat{\rho}_*^2 -\hat{\rho}_0^2\right) \\   -|\tilde V_*| \hat{ \rho}_* \int_{in} \frac{d^3x}{\mathcal V} (\psi({\bf x}) -\psi_0) \ ,
\end{multline}
where the scaled variables are defined by $\tilde \Omega \equiv \Omega/\hat V_0$, $\tilde {V}_* \equiv \hat{V}_*/\hat{V}_0$, and $\tilde \mu \equiv \mu/\hat{V}_0$, and the integral is carried out \textit{only inside the clusters}.

We compute $\tilde \Omega$ from numerical calculations of $\hat \rho_0$ and $\hat \rho_*$ in terms of the level-set parameter for candidate morphologies.
Here, we are assisted by the fact that $\rho({\bf x})$ is nearly $\rho_m$ in the cluster interior, where $\psi > \psi_0$, and is nearly 0 between clusters, when $\psi < \psi_0$.  Due to the non-linear relationship implied by eq. (\ref{rho}), this step-wise density modulation occurs at low-temperature even when $\phi(\xv)$ is dominated by a single mode.  For lamellar phases, we can compute the dependence of structural quantities on $\psi_0$ analytically. We find $\hat \rho_0 =\cos^{-1}(\psi_0)/ \pi$ and $\hat \rho_*=\sin[\cos^{-1}(\psi_0)]/\pi$.

\begin{figure}[t] \centering \includegraphics[angle=0, scale=0.35]{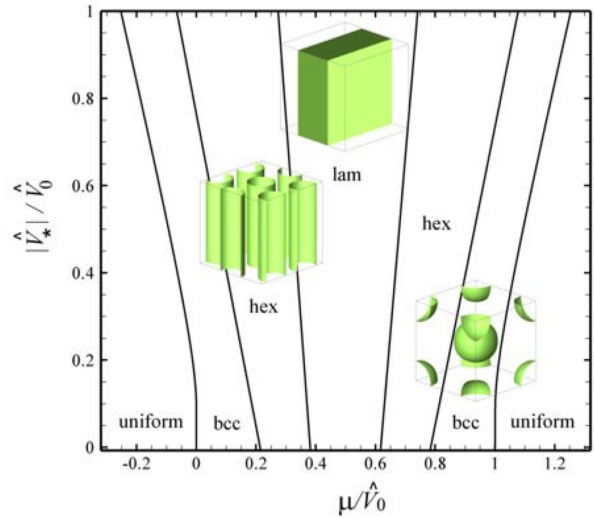} \caption{(Color online) Zero
temperature phase diagram in strong segregation limit. Lamellar, hexagonal-columnar, bcc and their inverse phases (from
the holes) are found as stable structures.} \label{SSTphase} \end{figure}

We construct the phase diagram as a function of $(|\tilde {V}_*|, \tilde \mu)$ comparing the computed values of grand potential from the candidate periodic phases as well as from the uniform state with $\hat{\rho}_*=0$.   Our computed phase diagram is shown  for the range $|\tilde{V}_*|<1$ in Fig.~\ref{SSTphase}.  We find that lamellar, hexagonal-columnar, bcc phases, and their inverse phases are stable phases in the strong-segregation limit.  All transitions between phases of different symmetry are first-order.  Note that the diagram is symmetric with respect to $ \mu /\hat{V}_0=1/2$, which arises from the ``particle-hole" symmetry of the lattice Hamiltonian.  The relation Eq.~(\ref{mu}) returns $\tilde \mu'=1-\tilde \mu$ for the holes  by $\hat\rho'_0=1-\hat \rho_0$, $\hat\rho'_*=-\hat\rho_*$ and $\psi_0=-\psi_0$.

\begin{figure}[t] \centering \includegraphics[angle=-90, scale=0.35]{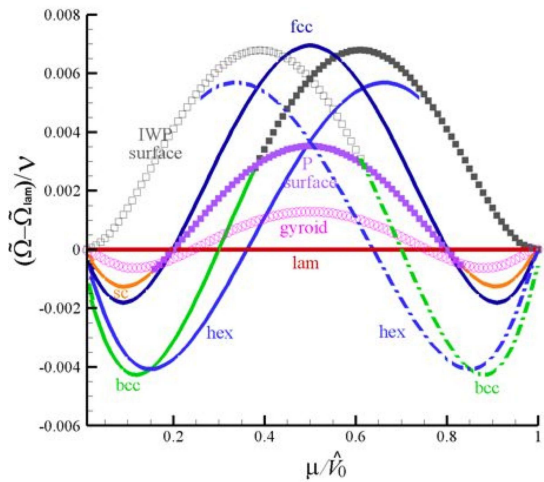} \caption{(Color online) The grand
potential differences from the lamellar phase for various candidate structures as a function of the chemical potential
$\mu/\hat{V}_0$. } \label{grand} \end{figure}

In Fig.~\ref{grand}, we also plot the grand potential energy differences from lamellar phase
as a function of $\tilde \mu$ for various structures along the line of $|\tilde {V}_*|=0.2$.
Interestingly, the gyroid phase is the second most stable phases in the range of $0.4 \lesssim \tilde \mu \lesssim 0.6$ and the most stable among the bicontinuous phases by far. It is more stable than the lamellar phase at very small and large $\tilde \mu$, though it is pre-empted by the hexagonal phase.

Though our model is unable to distinguish between structures that differ only by how the hard cores pack, our phase diagram is in qualitative agreement with simulations and a direct minimization of the repulsive-shoulder energy that allows lattice packing of cores to adjust \cite{kahl3}. Moreover, our analysis suggests that this phase diagram is universal among particle interactions whose Fourier transform has a sharp, negative peak.

\section{Finite-temperature corrections}\label{sec:finiteT}

At finite temperature, but still in the mean-field regime, the clustering thermodynamics are modified by the corrections arising from the finite width of the interface between the high and low density regions. This interfacial width is given by $\Delta = 1/(\alpha |\nabla \psi|)$ and depends on both the large parameter $\alpha$ and the interfacial geometry. This interfacial region at the boundary of clusters, is the primary source of particle fluctuations in the low temperature limit, giving rise to entropic contributions to the zero-temperature theory described in the previous section.

To compute the corrections to the energy, we employ a Sommerfeld-like expansion. It is most convenient to compute our corrections from an alternate expression for the grand potential
\begin{multline}
\label{DFT}
\beta \Omega = \frac{\beta}{2} \int d^3 x d^3 x' \rho(\xv) V_{SS}(\xv-\xv') \rho(\xv') + \\
 \int d^3x S(\xv) -\beta \mu \hat \rho_0,
\end{multline}
where
\be
\label{entropy}
S(\xv) =\rho \ln(\rho/\rho_m) + (\rho_m -\rho)\ln(1-\rho/\rho_m)
\ee is local contribution to the entropy of a lattice gas. This form is equivalent to the standard density-functional theory and can be derived from the grand potential using the mean-field equations.

Both inside and outside of clusters where $\rho(\xv)$ is equal to $\rho_m$ and $0$ respectively, the local contribution to the entropy is $S(\xv)=0$.  In the low temperature regime $S(\xv)$ is only non-zero at the cluster interfaces where $\rho_m>\rho(\xv)>0$. We expand the grand potential in powers of $\psi({\bf x}) -\psi_0$ and use a Sommerfeld expansion for $\hat\rho_0$ and $\hat\rho_*$.
To lowest order  in $\tilde T$ we have the low temperature correction to the grand potential Eq.~(\ref{DFT})
\be
\frac{ \delta \tilde \Omega }{{\cal V}} = \hat \rho_0 \delta \hat \rho_0 -m |\tilde V_*|\hat \rho_* \delta \hat\rho_*
+\tilde{T} \delta S
-\tilde \mu \delta \hat \rho_0 - \hat \rho_0 \delta \tilde \mu  ,
\ee
where
\bea
\nonumber
\delta \hat \rho_0 &=& \frac{\pi^2}{6 \alpha^2} \frac{d^2 \hat \rho_0}{d \psi_0^2} \\ \nonumber
\delta \hat \rho_* & =&\frac{\pi^2}{6 \alpha^2} \frac{d^2 \hat \rho_*}{d \psi_0^2} \\ \nonumber
\delta \tilde \mu &= &\delta \hat \rho_0 -|\tilde V_*|(\delta \hat \rho_*) \psi_0 \\
\delta S &=&  \int \frac{dA}{\mathcal V} |\nabla \psi|^{-1}
=\frac{\pi^2}{3 \alpha} \frac{d \hat \rho_0}{d \psi_0}  \ ,
\eea
and $\tilde{T} = k_B T/\hat{V}_0$.  Note that all properties about the low-temperature fluctuations is encoded by the level-set geometry, by first or second derivatives of $\hat\rho_0 $ and $\hat\rho_*$ with respect to the level-set parameter $\psi_0$. This also allows us to assign geometrical meanings to these terms. For example, the entropy correction $\delta S$ is proportional to the interfacial \textit{volume}, roughly $\Delta \int dA$. When $|\nabla \psi|= {\rm const.} $, implying a constant interfacial width, this correction simply scales like the interfacial area. Similarly, $d^2 \hat{\rho}_0/d\psi_0^2$ can be associated with the curvature of the interface weighted by the interfacial width.  Since $\delta S$ represents the entropy of enhanced core fluctuations at the interface, its contribution to the grand potential is always negative.  This suggests an entropic preference for cluster morphologies with a high volumetric surface area, a negative surface energy.  This tendency for large surface area structures competes with the terms derived from soft-shoulder interactions already included within strong-segregation theory.

\begin{figure}[t] \centering \includegraphics[angle=-90, scale=0.35]{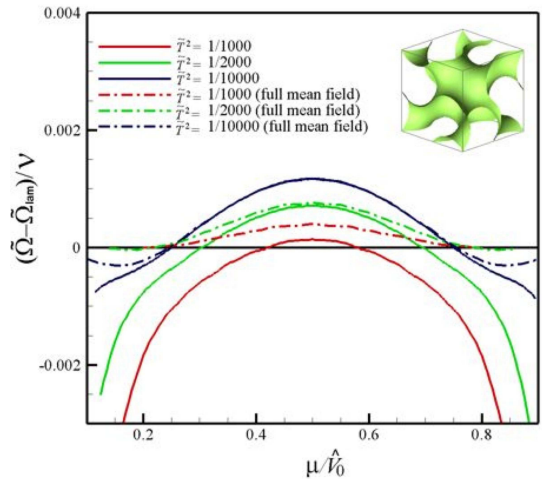} \caption{(Color online) Grand
potentials for the gyroid phase as temperature increases, $\tilde T^2=1/10000, 1/2000, 1/1000$ at $|\tilde {V}_*|=0.2$.   The
calculation from the lowest order corrections (solid lines) are compared to the full mean-field calculation (dashed lines). Note that the finite-temperature expansion becomes less accurate both as $\tilde T$ is increased and $\mu/\hat V_0$ approaches 0 and 1 (vanishing cluster width).} \label{gyroid}
\end{figure}

\begin{figure}[t] \centering \includegraphics[angle=-90, scale=0.35]{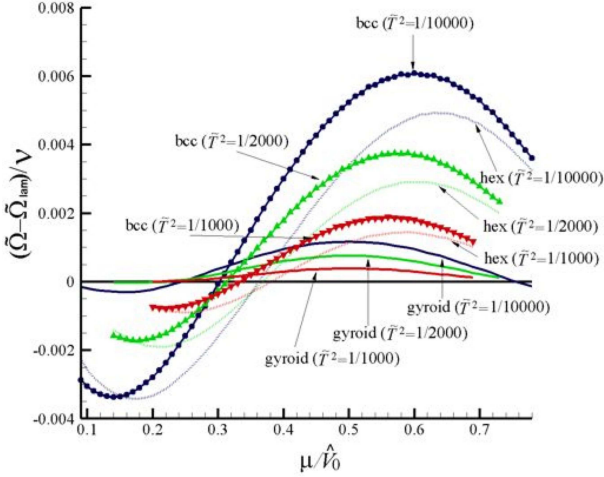} \caption{(Color online) The comparison
of grand potential energy among lamellar, hexagonal-columnar, and bcc phases from the full mean-field calculation for
temperature $\tilde{T}^2 =1/10000, 1/2000, 1/1000$ at $|\tilde {V}_*|=0.2$.
} \label{full} \end{figure}

As an illustration of the thermodynamics of low-temperature corrections described above, we consider the analysis at $\tilde \mu=1/2$ (or equivalently, $\hat \rho_0=\rho_m/2$, or $\psi_0=0$),
where the lamellar and gyroid phases are shown to be
the most and second most stable structures within strong segregation theory.  For both structures, $\psi_0=0$ is a point of inflection for the density so $\delta \hat \rho_0 =0$.   The contribution to finite temperacture corrections only arises from $\delta S$ and $\delta \hat \rho_*$.
In this case of the fixed density, we generally consider $\delta \hat \rho_*$ as to be negative
because the system becomes more homogenous as $T$ increases and the interface between core and cluster broadens.  At $|\tilde {V}_*|=0.2$,
the corrections to the grand potential per unit volume ($\lambda_*^3$) from $\hat \rho_*$, $-m |\tilde V_*|\hat \rho_* \delta \hat\rho_*$,
are evaluated as $\delta U_{lam} =4.1123 ~ \tilde{T}^2$ and $\delta U_{gyr}= 5.2592 ~  \tilde{T}^2$, for lamellar and gyroid phases, respectively.  Although the interaction energy of the gyroid phase increases more rapidly with temperature than that of the lamellar phase, we find that the mixing entropy correction plays a more dominant role in the overall grand potential energy change (per unit volume), giving $\delta S_{lam} =-8.2247 ~  \tilde{T}^2$ and $\delta S_{gyr}=-10.5197~ \tilde{T}^2$, for lamellar and gyroid phases, respectively.  The surface area of gyroid phases significantly lowers the grand potential
and thus become more stable at finite temperature relative to the zero temperature limit.
We plot the grand potential changes for the gyroid phases
as temperature increases ($\tilde{T}^2=1/ 10000$, $1/5000$ and $1/2000$) in Fig.~\ref{gyroid}.
Due to the invariance of the mixing entropy $S$  under inversion ($\rho \rightarrow \rho_m -\rho$),
the contribution of temperature corrections is symmetric with respect to $\rho=1/2$.

This lowest order calculation is expected to break down where the interfacial width $\Delta$ becomes larger than the actual cluster size of order $\lambda_*$.  This can occur either when temperature becomes large, or when $\hat{\rho}_*$ becomes sufficiently small.  This latter condition is always met as the density of particles or holes goes to zero at the boundaries of the segregation regime.  Therefore, the finite-temperature expansion first breaks down as $\tilde{\mu} \to 0$ or $\tilde{\mu} \to 1$.
To more carefully test the validity of the low-temperature expansion we also perform the full mean-field calculation by numerically solving the nonlinear self-consistent equations.  The results are displayed along with the lowest order calculations in Fig.~\ref{gyroid}.  In order to assess the equilibrium stability of the bicontinuous gyroid in the finite temperature phase diagram, we plot the grand potential for hexagonal and bcc phases, which are the stable phases in the strong segregation regime.
The full calculation results for lamellar, gyroid, hexagonal, and bcc phases
are presented in Fig.~\ref{full} for $\tilde{T}^2=1/10000$, $1/5000$, $1/2000$ and $|\tilde{V}_*|=0.2$.  Despite the trend suggested by the lowest-order finite temperature corrections, we find that the gyroid structure is not stable at finite temperature and the topology of the mean-field phase diagram is not qualitatively changed in the low temperature regime due to the presence of finite-temperature corrections.

\section{Weak-segregation limit}\label{sec:highT}

Intrigued by near stability of the gyroid phase in the finite-temperature expansion, we pursue the mean-field clustering phase behavior near to the order-disorder transition, where the density is nearly constant with only small amplitude modulations.  Here, we follow Liebler's approach to thermodynamics of block copolymer melts ~\cite{leibler} near the critical point to derive a Landau theory expansion for thermodynamic potential in terms of the lattice mean-field theory of clustering.   As argued by Alexander and McTague, the form of the expansion of the free energy in terms of periodic density modulations near the liquid-to-solid phase transition takes on a rather generic form~\cite{alexander_mctague}.  Our task is to derive expressions for the coefficients of this expansion in terms of the thermodynamic parameters of the cluster model:  $\beta$, $\hat{V}_0$, $\hat{V}_*$ and $\mu$.  This is most concisely performed by first using the density functional theory for the Helmholtz free energy (fixed density), which is related to the mean-field theory of the action in equation \ref{action}.
This yields
\begin{equation}
\beta F = \int d^3x S (\xv)  +\frac{\beta}{2} \int d^3x'~\rho({\bf x}) V_{SS} ({\bf x}-{\bf x}') \rho({\bf x}') ,
\end{equation}
where $S(\xv)$ is the lattice-gas expression for the local contribution to the entropy, eq. (\ref{entropy}).
We write $\rho({\bf x}) = \rho_0 + \delta \rho({\bf x})$, where $\int d^3x~\delta \rho({\bf x}) = 0$ and $\rho_0$ is constant.  To Legendre transform to the grand canonical ensemble, we add a term $- {\cal V} \mu \rho_0$ and minimize with respect to $\rho_0$ for a fixed density variation.  Expanding to quadratic order in $\delta \rho$, we find the grand potential
\begin{eqnarray}
\label{landau}
\frac{\beta \Omega}{\mathcal{V}} &=& \frac{1}{2} \sum_{\bf q} \Gamma_2({\bf q}) | \delta \hat{\rho} ({\bf q} )|^2 + \frac{\gamma}{3!} \int \frac{d^3x}{\mathcal{V}}~\delta \rho^3({\bf x})\\
& & + \frac{\lambda}{4!} \int \frac{d^3x}{\mathcal{V}}~\delta \rho^4({\bf x}) - \frac{\delta}{4} \left( \int \frac{d^3x}{\mathcal{V}} \delta \rho^2({\bf x}) \right)^2.\nonumber
\end{eqnarray}
The coefficients are defined in terms of the function $A(\rho) = \rho \ln (\rho/\rho_m) - (\rho_m- \rho) \ln (1 - \rho/\rho_m )$.  The mean value of density, $\bar{\rho}_0$, is determined by the solution to the equation
\begin{equation}
\hat{V}_0 \bar{\rho}_0 + A^{(1)} (\bar{\rho}_0)  = \beta \mu,
\end{equation}
where  $A^{(n)} = d^n A/d \rho^n$. This allows us to write the coefficients as
\begin{eqnarray}
\Gamma_2({\bf q}) &=& A^{(2)}( \bar{\rho}_0) + \hat{V} (\qv) \nonumber\\
\gamma &=& A^{(3)}(\bar{\rho}_0)\nonumber\\
\lambda &=& A^{(4)}(\bar{\rho}_0)\\
\delta &=& \frac{[A^{(3)}(\bar{\rho}_0)]^2}{2 (\hat{V}_0 + A^{(2)}(\bar{\rho}_0)}\nonumber
\end{eqnarray}
Note that the final, non-local quartic term in the Landau expansion of grand potential does not appear in original analysis of Alexander and McTague, nor within the mean-field treatment of diblock copolymer melts~\cite{leibler}.

At this point we now consider the grand potential for various periodic structures.  As before, the function of $\hat{V}(\qv)$ is to select structures with a periodic modulation of density at $|\qv|=q^*$, to minimize the values of the quadratic coupling of the density, $\Gamma_2(\qv) |\delta \hat{\rho}(\qv)|^2$.  Thus, we consider density variations which, like $\psi(\xv)$, are plane wave superpositions of a given symmetry group.  Following a similar analysis for copolymer melts~\cite{ranjan_morse}, for a given symmetry the density modulation can be expanded in terms of the basis functions
\begin{equation}
\delta \rho({\bf x}) = \sum_i \Psi_i \phi_i({\bf x})
\end{equation}
where $\Psi_i$ is an amplitude (not related to level-set parameter) and $\phi_i({\bf x})$ is given by a superposition of plane waves
\be
\phi_i({\bf x}) = \sqrt{\frac{2}{n_i}} \sum_{{\bf G}_i} c_{\Gv_i} e^{i {\bf G}_i \cdot {\bf x}}
\ee.
Here, ${\bf G}_i$ are a family of reciprocal basis vectors corresponding to the symmetry of a candidate structure and the $c_{{\bf G}_i}$ are phase factors satisfying $|c_{{\bf G}_i}| = 1$. We again consider only modes $|{\bf G}_i| = q^*$.

\begin{figure}[t] \centering \includegraphics[angle=0, scale=0.5]{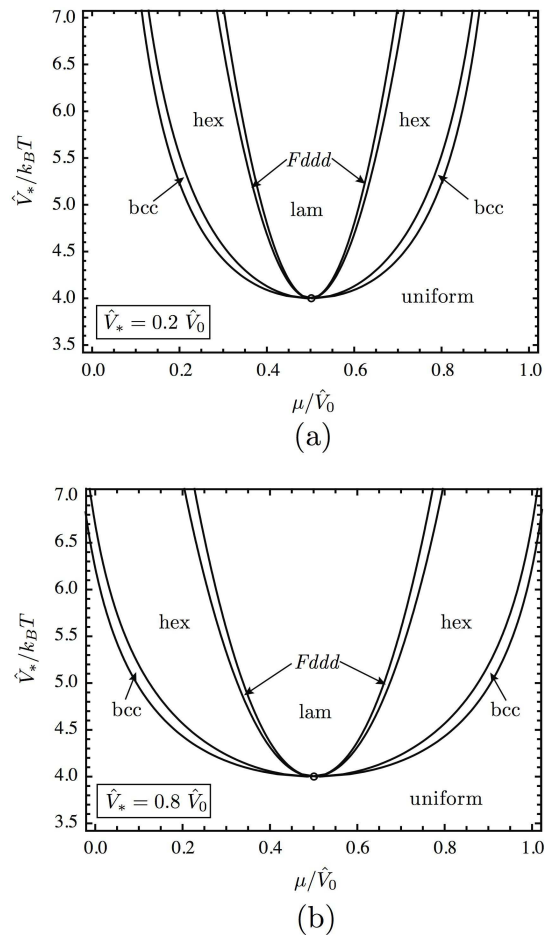} \caption{The mean-field phase diagrams of isotropic, clustering particles in the limit of weak-segregation.  The phase diagrams for two soft-shoulder potentials are shown:  $|\hat{V}_*|=0.2 \hat{V}_0$ in (a) and $|\hat{V}_*|=0.8 \hat{V}_0$ in (b).  All phase transitions are first-order except at the critical point (dark circle) where the system may transition continuously from the disordered state to the lamellar phase.} \label{weak} \end{figure}

It was the original insight of Alexander and McTague, that the mean-field behavior of the liquid-to-solid transition is dominated by structures that contain a large number of triplets for which $\Gv_i+\Gv_j+\Gv_k=0$.  This condition allows the cubic term in eq. (\ref{landau}) to be maximally negative, lowering the free energy of certain periodic structures..  It is this geometrical feature of the reciprocal lattice of the bcc structure (8 triplets) which generically favors this structure in the mean-field theory of the liquid-to-solid transition ~\cite{alexander_mctague} as well as in the phase diagram of weakly segregated block copolymer melts ~\cite{leibler}.

More recently it was discovered by Morse and coworkers~\cite{tyler_morse, ranjan_morse} that there was a non-cubic bicontinuous structure, the orthorhombic {\it Fddd} structure, that was both unimodal and contained an abundance of fundamental mode triplets that add to zero.   Structurally, this $Fddd$ structure is related to an orthorhombic distortion of the cubic {\it double-gyroid} structure with $Ia\bar{3}d$ symmetry which enforces the condition that all 14 wavevectors to have the same length.  The $Fddd$ phase is constructed from three families of modes: ${\bf G}_1 =2 \pi [\pm\sqrt{3} \hat{x}/2 \pm \sqrt{3} \hat{y}/4\pm\hat{z}/4]/\lambda_*$, ${\bf G}_2 = 2 \pi [\pm\sqrt{3}  \hat{y}/2  \pm \hat{z}/2]/\lambda_*$ and ${\bf G}_3 = 2 \pi [\pm  \hat{z}]/\lambda_*$. Therefore, for this structure we also have three independent amplitudes $\Psi_i$ to consider.

Using the above Landau expansion, we derive the following expressions for the grand potential
\begin{eqnarray}
\omega_{lam} &=& \Gamma_2(q^*) \Psi^2 + \left(\lambda/4 - \delta \right) \Psi^4\nonumber\\
\omega_{hex} &=& \Gamma_2(q^*) \Psi^2 - \frac{2}{3 \sqrt{3}} \Psi^3 + \left( \frac{5}{12} \lambda - \delta\right) \Psi^4\\
\omega_{bcc} &=& \Gamma_2(q^*) \Psi^2 - \frac{4}{3 \sqrt{6}} \gamma \Psi^3 + \left(\frac{5}{8} \lambda - \delta\right) \Psi^4\nonumber\\
\omega_{Fddd} &=& \Gamma_2(q^*) \left(\Psi_1^2 + \Psi_2^2 + \Psi_3^2\right)\nonumber\\
& & - \gamma \left(\frac{1}{\sqrt{2}} \Psi_1^2 \Psi_2 + \Psi_2^2 \Psi_3\right)\nonumber\\
& & + \lambda \Big(\frac{5}{16} \Psi_1^4 + \frac{3}{8} \Psi_2^4 + \frac{1}{4} \Psi_3^4\nonumber\\
& & + \Psi_1^2 \Psi_2^2+\Psi_2^2 \Psi_3^2 + \Psi_1^2 \Psi_3^2 + \frac{1}{\sqrt{2}} \Psi_1^2 \Psi_2 \Psi_3\Big)\nonumber\\
& & - \delta \left(\Psi_1^2 + \Psi_2^2 +\Psi_3^2\right)^2\nonumber.
\end{eqnarray}
The mean-field phase diagram near the critical point is computed by minimizing over $\Psi_i$ for each candidate structure and comparing values to find the structure which minimizes the potential.

For two values of $\hat{V}_*/\hat{V}_0$ we show the computed phase diagrams in Fig. \ref{weak}.  All phase transitions are predicted to be first-order except at the critical point $\beta \hat{V}_* = 4$ and $\mu = \hat{V}_0/2$ where the system is predicted to pass from the uniform, disordered state to the lamellar phase via a continuous, second-order phase transition.  Interestingly, near the onset of order the the same cluster lattices are stable: lamellar, hexagonal columnar and bcc lattice of spherical clusters.  More surprising is the fact the mean-field theory predicts a narrow window of a stable bicontinuous morphology, between the lamellar and columnar morphologies.

Thus, we predict a mean-field phase diagram for cluster particles that is on the whole rather similar to that was originally predicted by Leibler \cite{leibler}--and updated by Ranjan and Morse~\cite{ranjan_morse}--for diblock copolymer melts.  In that system, beyond the weak-segregation theory, it becomes crucial to consider periodic structures with density modulations for which $|\qv| \neq q^*$.  In particular, this is a critical ingredient for establishing that the double-gyroid structure, rather than the orthorhombic $Fddd$ morphology, is by far the most prominent equilibrium bicontinuous structure in copolymer melts \cite{olmsted_milner,schick}.  Indeed, it is quite reasonable to expect that a more realistic model of a soft-shoulder potential that includes values of $\hat{V}(\qv)$ for ``off-peak" wavevectors will predict the stable bicontinuous structure to be the cubic double-gyroid morphology.  However, this complication is beyond the scope of our present analysis.  This shortcoming notwithstanding, our present analysis is sufficient to conclude--at least within some very narrow region of phase space in the neighborhood of the critical point--that there is a stable bicontinuous morphology within the mean-field phase diagram of isotropic, clustering particles.

Finally, it is important to note that our predictions have focused strictly within the limit of mean-field.  It is well-known that near to the order-disorder transition fluctuations of the order parameter--long-wavelength excitations of the ordered structures--as well as fluctuations within the disordered play an important role in determining phase behavior~\cite{brazovskii}.  Within the context of block copolymer melts, for example, these fluctuations shift the thermodynamic onset of periodic order to slightly lower temperatures and generally promote the stability of the lamellar morphology with its abundance of soft-modes~\cite{fredrickson_helfand, muthukumar}.  It would be straightforward to adapt those theoretical methods to the analyses of clustering mesophases, and expect the effect of thermal fluctuations to play a similar role in the present context.

\section{Discussion}\label{sec:discussion}

We have presented calculations of the three-dimensional phase diagram of a lattice gas of particles interacting through a soft corona. When the interaction potential displays a Fourier transform with a sharp, negative peak, we find a \textit{universal} phase diagram in both the strong-segregation (low temperature) and weak-segregation (near critical) limits. These phase diagrams mirror those found in many soft systems, most prominently for diblock copolymer melts.  In the limit of weak-segregation, this correspondence becomes more concrete as the mean-field model describing the onset of periodic order has the same generic form for both systems.  Though our lattice model is unable to distinguish between the different core packings leading to these complex ground states, our results corroborate generally numerical calculations by Pauschenwein and Kahl~\cite{kahl2,kahl3} of the ground state of particles with a particular soft-shoulder potential, the step-function coronas described in Eq. (\ref{eq:step}).  A notable divergence of these analyses occurs at very high particle densities, or large chemical potential.  Our results demonstrate a generic equivalence--even at low temperature--between two- and three-dimensionally modulated cluster morphologies and their inverse structures [see Fig. (\ref{SSTphase})].  These inverse morphologies are absent from the high-density predictions of refs. \cite{kahl2} and \cite{kahl3}, but appear in finite-temperature, Monte Carlo simulations of step-shoulder potentials in two-dimensions~\cite{glaser_et_al}.

Moreover, we are able to obtain systematic corrections to the low temperature behavior and find that these act as an effective negative surface tension -- thermal fluctuations prefer to create interface. This occurs because the entropy is concentrated along the cluster interfaces. This surface tension decreases the gyroid free energy but does not, apparently, stabilize it at temperatures in which our low temperature expansion is valid. The low energy of the gyroid compared to other bicontinuous phases can be attributed to the fact that the trigonometric approximation to the gyroid has a smaller mean curvature, and is therefore an extremum of the area, compared to other symmetry structures. This scenario is similar to a ``packing frustration" which occurs in block copolymer morphologies~\cite{matsen_bates, grason}. In that case, the propensity for an extremal interface between two immiscible blocks is frustrated by the stretching of the polymers. Apparently, the double gyroid is the extremal surface in which the polymers are the most uniform. In the case of clustering particles, this three-dimensional uniformity is related to the presence of a single characteristic length scale in the interaction potential and the entropy frustrates this energetic contribution.

For clustering interactions in which a single length scale dominates, we can also turn these results around to give a geometrical interpretation of cluster morphologies in the strong-segregation regime. On the one hand, energetics (roughly) minimize the number of corona overlaps while, on the other hand, entropy maximizes interfacial area.

In the weak-segregation regime, coarse-grained approaches to colloidal assembly with competing interactions of various levels of sophistication have predicted universal phase diagrams similar to ours \cite{seul95,gelbart,ciach}. We are aware of one prediction in this regime of a possible bicontinuous phase with cubic symmetry \cite{ciach}. Our results give this universality a compelling geometrical interpretation at \textit{low} temperatures in terms of minimizing the number of overlaps while simultaneously maximizing interfacial area under appropriate constraints. The relatively low energy of the gyroid phase suggests that the possibility that such phases could be observed as a metastable structure, and could even become stable for wider negative peaks that allow additional nearby length scales.

What lessons can we glean from this analysis for ``designing" potentials to exploit the complex phase behavior of clustering particles?  Foremost, we note that the morphological characteristics are controlled by the shape of the potential in Fourier space. Additional morphologies are likely to be found by tuning potentials to select one or more characteristic length scales. Future work will explore the role of multiple, negative peaks and their widths in the Fourier transform. In particular, such multiple peaks might stabilize unusual phases, including quasicrystals \cite{lifshitz} or an A15 lattice of clusters.

\acknowledgements{It is pleasure to thank P. Ziherl for a careful reading of this manuscript.  HS and GMG acknowledge support from the NSF center grants DMR-0829596 and CMMI-0531171.}

\end{document}